\documentclass{elsart}

\usepackage[round,colon,authoryear]{natbib}

\usepackage{epsfig}

\usepackage{amssymb}

\begin{document}

\begin{frontmatter}

\title{Star-Forming AGN Host Galaxies}

\author{Peter Barthel}

\address{Kapteyn Astronomical Institute, P.O. Box 800, 
 9700 AV~~Groningen, The~Netherlands}

\ead{pdb@astro.rug.nl}

\begin{abstract}
The symbiosis of nuclear activity and star-formation in galaxies, as 
manifested in their spectral energy distributions (SEDs) is reviewed.
Attention is drawn to an Hertzsprung-Russell diagram-equivalent for such 
objects, as well as to the importance of the SEDs in cosmological context.
\end{abstract}

\begin{keyword}
galaxies: active \sep galaxies: starburst \sep quasars: infrared 
\end{keyword}

\end{frontmatter}

\section{Introduction}

\def\putplot#1#2#3#4#5#6#7{\begin{centering} \leavevmode
\vbox to#2{\rule{0pt}{#2}}
\includegraphics{#1}
\end{centering}}

Let me start this review (the first paper at this workshop) by showing
you a picture of good-old IRAS.  It is fair to say that IRAS added a new
dimension to the parameter space of galaxies: dust emission.  This
dimension is the very reason for our get-together! Not only has the
highly diverse dust content of normal and active galaxies become
apparent, but a new class of very dusty galaxies has also emerged: the
UltraLuminous Infrared Galaxies (ULIRGs).  This review will primarily
discuss the active galaxies among the former group, but also deal with
the latter group, in their relation to the former. 

A key issue in contemporary astrophysics is the cosmologically evolving
star-formation activity.  Nuclear activity, taken here to mean accretion
driven energy generation, and circumnuclear star-formation activity in
galaxies, are sometimes seen to go hand in hand (e.g., NGC\,1068:
Schinnerer et al.  2000; Mkn\,231: Taylor et al.  1999).  The question
arises whether the resemblance of the cosmic star-formation history
diagram (Madau et al.  1996) and the plot of evolving QSO space density
(in particular the $z \approx 2.5$ peak; e.g., Shaver et al.  1996) is
purely coincidental.  In this review I wish to convey the message that
AGN host galaxies offer excellent opportunities to study the history of
star-formation over a substantial cosmic time span, and I will describe
the usefulness and the unique character of the global Spectral Energy
Distribution (SED) of these objects in such studies.  Indeed, as judged
from these SEDs, a substantial fraction of (nearby) QSO host galaxies
display vigorous star-formation.  I will draw attention to a
far-infrared AGN Hertzsprung-Russell diagram that may prove useful in
the study of the interconnection between active galaxies and ULIRGs. 

\section{Dust in normal and active galaxies}

The FIR SED of a normal galaxy is commonly believed to consist of a cold
(T$\sim$15\,K) component and a warmer (T$\sim$20--60\,K) component
(e.g., Helou 2000).  The former is the well known large scale cirrus
component, being energized by the diffuse interstellar radiation field
of the, primarily, old stellar population.  The latter originates from
dust in star-forming regions, i.e., regions containing massive young
stars which provide a strong, dust-heating UV photon field (and
moreover, yield supernovae).  Depending on the relative strength of
these two components a galaxy will display a cold, intermediate
temperature, or warm FIR SED.  While passively evolving spirals have a
steeply rising SED from 60\,$\mu$m to 100\,$\mu$m,
with\footnote{$S_{\nu} \propto {\nu}^{\alpha}$ adopted} average
$\alpha_{60}^{100} \sim -1.5$ (or equivalently log\,$S$(60)/$S$(100)
$\sim -0.3$) and average $\alpha_{25}^{60} \sim -2.5$, starburst
galaxies such as M\,82 and Arp\,220 are warmer, having
$\alpha_{60}^{100} \sim 0$ (and $\alpha_{25}^{60}$ in the range $-2$ to
$-3$).  The strength of the warm dust component is correlated with the
optically thin synchrotron radio emission -- this is the well known
radio-FIR correlation (e.g., Condon 1992). 

Active galaxies display still warmer dust, drawing additional heating
input from the hard continuum radiation of the active nucleus.  Seyfert
galaxies for instance display warm FIR SEDs, and the FIR warmth as
quantified by the $\alpha_{25}^{60}$ index indeed appears to be a useful
AGN selection criterion: De Grijp et al.  (1985) found a high fraction
of Seyferts among galaxies having $\alpha_{25}^{60}$ flatter
than $-1.5$.  Early on in the IRAS survey the nearby broad-line
radio galaxy 3C\,390.3 was suspected to have even warmer dust: Miley et al. 
(1984) reported a pronounced 25\,$\mu$m peak in this object. Van Bemmel
(these Proceedings) deals with the far-infrared SEDs of the BLRG class,
suggesting that in fact there is not an excess of hot dust but a
deficit of cool dust in these 25\,$\mu$m peakers.

While ground-breaking work on FIR SEDs in galaxies was carried out by
IRAS, ISO has put this research on a more quantitative basis.  For
instance, Rodr\'{\i}guez Espinosa et al.  (1996) and P\'erez Garc\'{\i}a
et al.  (1998) used ISO data to further constrain the multi-temperature
dust model for Seyfert galaxies.  From detailed measurements of the FIR
SEDs they infer the presence of warm dust (T$\sim$150\,K), presumably
heated by AGN, and cooler dust (T$\sim$45\,K), presumably heated by
young stars.  Also, the cold cirrus component could be identified in a
number of cases.  Subsequent optical imaging (Rodr\'{\i}guez Espinosa \&
P\'erez Garc\'{\i}a 1997) demonstrated that indeed, the warm and cool
dust are respectively AGN and host galaxy related.  Bimodal dust
temperature distributions were also measured in the interacting Antennae
galaxies NGC\,4038,\,39 and the ultraluminous infrared galaxies
NGC\,6240 and Arp\,220 (Klaas et al.  1997).  In these objects the warm
dust is believed to be powered by starburst activity, the relative
strength of the warm component w.r.t.  the cool component being governed
by extinction within the system. 

While the star-formation related dust cannot be easily linked to the
color and luminosity of the host galaxy (e.g., Helou 2000), such an
astrophysical connection must be present.  Bregman et al.  (1998) for
instance found that FIR-bright early type galaxies are more luminous in
H$\alpha$, H\,I and CO lines as compared to their normal counterparts. 
In the same vein, Clements (2000) showed that among radio-quiet QSOs the
ones having cool colors (hence substantial cool dust besides the AGN
related hot dust) stood out with a higher FIR luminosity and higher
incidence of distorted optical host morphology.  The latter is generally
taken to imply enhanced star-formation (e.g., Mazzarella et al.  1991). 

\section{Dust in quasars and radio galaxies}

As for the more powerful AGN in the distant universe, it should be first
noted that IRAS detected primarily radio-loud objects, with a dominant
contribution from flat (radio-)spectrum objects.  This calls for a
careful examination of the SEDs of radio-loud quasars and radio
galaxies, taking their detailed radio characteristics into account.  The
optical--FIR--radio SED of the well known radio galaxy Cygnus~A
(3C\,405), as compiled by Ilse van Bemmel, is shown in Fig.~1, together
with a model fit of optically thin black bodies.  The infrared spectral
indices of Cyg~A are: $\alpha_{60}^{100} = +1.1$ and $\alpha_{25}^{60} =
-1.3$. 

\begin{figure}[h]
\putplot{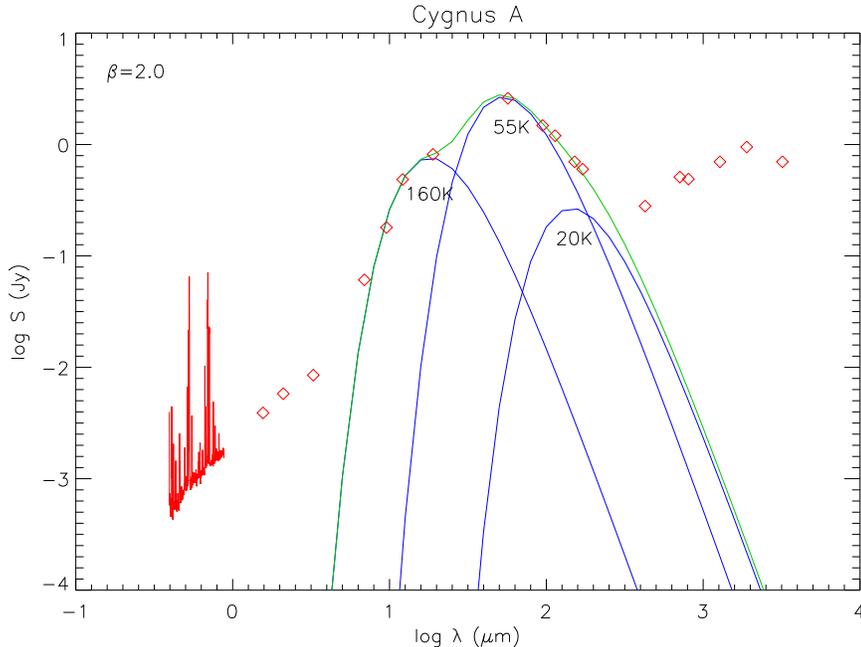}{7.8cm}{0}{70}{70}{0}{-270}
\caption{
The optical--NIR--FIR--submm--radio SED for Cygnus~A.  Ground-based
optical, near-IR and radio data are combined with ISO photometry.  The
sub-mm, mm, and cm points reflect the core emission, obviously excluding
the extended radio lobes.}
\end{figure}

Consistent with the FIR SEDs of normal and Seyfert galaxies, we see
cold, cool and warm dust in Cyg~A.  The pronounced cool (T=55\,K) dust
luminosity suggests a substantial level of star-formation in the Cyg~A
host galaxy.  The fact that the ratio of FIR and radio luminosity for
Cyg~A is substantially different from the general radio galaxy
population is nevertheless attributed to the anomalously high radio
luminosity of the object (Barthel \& Arnaud 1996).  As can be readily
seen, non-thermal FIR radiation, i.e., the short wavelength tail of the
nuclear continuum, has an almost negligible strength.  Combining clear
detections at about ten wavelengths, no other radio galaxy or quasar has
its infrared SED so well determined. 

As for the quasars measured by IRAS, very few distant radio-quiet QSOs
and only a handful of steep-spectrum 3CR QSRs were detected (Neugebauer
et al.  1986).  Early on, QSR 3C\,48 was recognized as an extraordinary
luminous FIR emitter (Neugebauer et al.  1985).  QSR 3C\,446 was also
found to be infrared luminous, but as nicely illustrated by its IRAS
measured variability (Neugebauer et al.  1986), nonthermal radiation
must be dominating the 3C\,446 SED, in line with its identification as a
Blazar.  Blazars were soon recognized as a class of their own, with
smooth albeit variable radio--FIR--optical continua (Edelson \& Malkan
1987, Impey \& Neugebauer 1988, Brown et al.  1989).  The IRAS radio
galaxy study by Golombek et al.  (1988) yielded a couple dozen
detections but only two objects beyond $z=0.2$.  However, despite the
limitations of these data, some interesting trends emerged for various
classes of powerful AGN.  For instance, Golombek et al.  found that the
radio galaxy FIR strength reflected their emission line strength. 
Heckman et al.  (1992, 1994), employing improved IRAS data analysis
routines, compiled infrared SEDs for ensembles of AGN.  They reported
that FR1 (edge-darkened) radio sources were about a factor of four less
luminous in the FIR as compared to FR2 (edge-brightened, hotspot)
sources, and they confirmed the presence of a hot 25\,$\mu$m dust
component in the class of broad-line radio galaxies.  They also reported
subtle differences in FIR output between FR2 radio galaxies and QSRs,
discordant with predictions of the simple orientation-based unification
model. 

The work of Hes et al.  (1995) increased the detection rate of distant
3C objects and provided a clear picture of the mechanisms involved. 
They confirmed the Heckman et al.  finding that quasars are somewhat
brighter in the far-infrared than radio galaxies, but pointed out that
beamed non-thermal radiation might be (at least in part) responsible for
this difference.  The 60\,$\mu$m strength being correlated with the
radio core strength, it is unlikely that the Rowan-Robinson (1995)
two-component dust model applies to radio-loud objects: the AGN strength
is probably also reflected in the long wavelength FIR emission. 
Hoekstra et al.  (1997) modeled FIR beaming, and quantified its effect
on the FIR SED, as a function of radio beaming.  Hes et al.  (1995) also
discovered that the probability of detecting a radio galaxy in the
far-infrared is enhanced by optical N-type morphology, in connection
with either broad lines or the combination of compact radio structure
and blue optical colors.  The latter fact was clearly underlined by
Willott et al.  (2000), who discovered that the redshift of the IRAS
detected compact radio galaxy 3C\,318 is 1.5 rather than 0.7: 3C\,318 is
the most distant 3C source detected by IRAS, harbouring a dust
luminosity of approximately $10^{14}\,L_{\odot}$.  The high FIR
luminosity of the QSR 3C\,48 could be understood along similar lines. 
The subgalactic size radio source is hosted by a galaxy merger
displaying luminous young star clusters and HII regions (Kirhakos et al. 
1999, Canalizo \& Stockton 2000b) as well as bright molecular gas
emission (Wink et al.  1997).  The FIR luminosity of 3C\,48 must be
connected to ongoing star-formation. 

Combining IRAS with cm and mm data, Van Bemmel et al.  (1998) further
quantified the level of FIR beaming in double-lobed quasars: they found
its magnitude to be modest, and not sufficient to explain the difference
with radio galaxies.  Using the PHOT instrument on board ISO Haas et al. 
(1998) confirmed and refined the picture as drawn by IRAS.  In
particular, the contributions of beamed non-thermal and presumed
aspect-independent thermal FIR emission could be isolated in the SED of
blazar 3C\,279, and the multicomponent nature of its dust emission
became apparent.  It will by now be clear why I started my examination
of the AGN FIR SED components with the radio-loud subclass: these offer
the advantage that their jet orientation w.r.t.  the observer is known
so that orientation dependence can be taken into account.  Focussing on
the unification issue, such was again done by Van Bemmel et al.  (2000),
examining ISO photometry of samples of quasars and radio galaxies. 
Apparent disagreement with predictions of the unification scheme may be
attributed to small anisotropy effects (but see also M\"uller et al.  in
these Proceedings). 

Polletta et al.  (2000) examined ISOPHOT measurements of the FIR SEDs of
quasars, dealing with various classes of radio-loud and radio-quiet
objects.  They present average quasar SEDs and conclude that AGN
powered far-infrared emission dominates the SED, but that it is
supplemented by starburst powered FIR emission.  The latter is observed
in radio-loud and radio-quiet QSOs, at comparable strength in both
classes.  Haas et al.  (2000; see also Haas' contribution in these
Proceedings) describe ISO measurements of the FIR SEDs of PG QSOs.  The
SEDs suggest strong dust emission, drawing power from both the AGN and
starburst activity. 

While in the case of radio-loud quasars and radio galaxies radio images
at arcsecond and milliarcsecond resolution irrefutably demonstrate that
the activity originates in a subparsec sized volume, such is not readily
apparent in radio-quiet objects.  Given the fact that radio-quiet QSOs
appear to follow the radio-FIR correlation for star-forming systems
(which include 30\,Doradus, irregular galaxies, spiral galaxies and
ultraluminous infrared galaxies) it has been proposed that QSOs may be
powered by starburst activity as well (e.g., Sopp \& Alexander 1991). 
Analyses by Rush et al.  (1996) and Giuricin et al.  (1996) have in the
meantime shown that excess radio emission is present in Seyfert
galaxies, calling for a non-thermal AGN.  The issue for the more distant
and powerful QSOs remains controversial.  I will return to this issue
later. 

In summary, the FIR SEDs of active galaxies are composed of cold ISM
dust emission, cool star-formation dust, warm AGN heated dust and
possibly a beamed nonthermal component.  Describing the FIR SEDS in
terms of discrete temperature components (besides the smooth synchrotron
continuum) is obviously a naive simplification, but so far seems to work
reasonably well. 

\begin{figure}[h]

\putplot{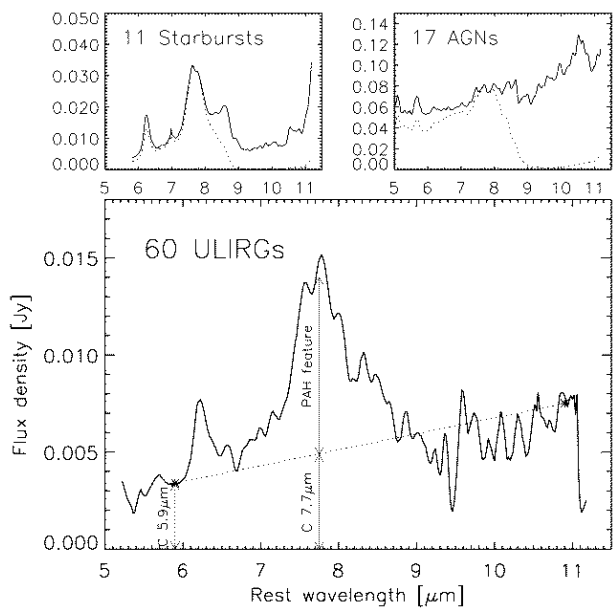}{8cm}{0}{60}{60}{0}{-140}

\caption{The relative strength of the 7.7$\mu$m PAH feature for the
 average (nearby) AGN, starburst and ultraluminous infrared galaxy, 
 from Lutz et al. (1998).}
\end{figure}

\section{Dust and gas in ULIRGs and QSOs}

We have seen in the previous sections that both AGN heated and
star-formation heated dust can be recognized in AGN FIR SEDs.  Prime
examples of systems undergoing vigorous star-formation are the ULIRGs
(e.g., Sanders \& Mirabel 1996).  Also given the fact that radio-quiet
QSOs and ULIRGs obey the same radio-FIR correlation (see for instance
Figures 3 and 4 in Colina \& P\'erez-Olea 1995), the question as to the
interconnection between AGN and starbursting ULIRGs is a long standing
one -- e.g., Sanders et al.  (1988).  At least for nearby objects, ISO
spectroscopy has indicated that the ionization stages of AGN and ULIRGs
are different: high ionization lines ([O\,IV], [Ne\,V]) appear in the
mid-infrared spectra of the former but not the latter class (Lutz et al. 
1996, Lutz et al.  1999).  Also the relative strength of the PAH
features appears to be a useful diagnostic -- see Fig.~2, taken from
Lutz et al.  (1998), which shows this PAH/continuum ratio for the
average AGN, starburst galaxy and ULIRG.  Spoon et al.  (2000) has
recently pointed out a caveat when using this diagnostic, and it should
also be noted that these diagnostics have not yet been applied to
luminous QSOs. 

\begin{figure}[h]

\putplot{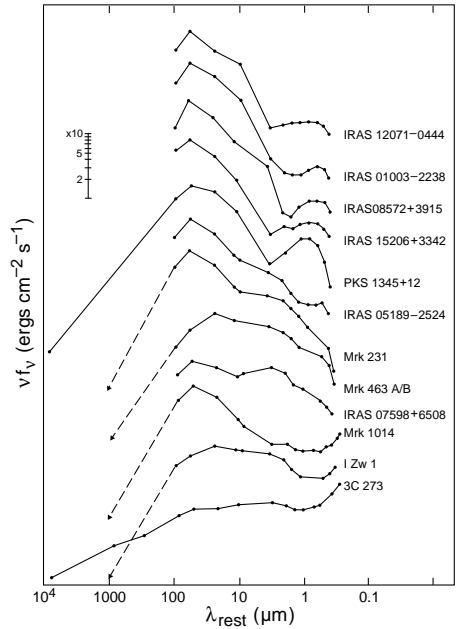}{8.5cm}{0}{80}{80}{50}{-40}

\caption{Radio--FIR--optical SEDs for warm IRAS sources
 (Sanders et al. 1988)}
\end{figure}

Section~2 already addressed the relative strength of warm and cool dust,
and specifically the FIR SED spectral slope $\alpha_{25}^{60}$, to
isolate AGN among star-forming galaxies.  The usefulness of this index
is nicely illustrated by Fig.~3, a compilation of SEDs of relatively
warm ultraluminous infrared sources, taken from Sanders et al.  (1988):
a range of spectral slopes is seen, from fairly steep to flat/inverted. 
Fig.~4 combines $\alpha_{25}^{60}$ with $\alpha_{60}^{100}$ data, for
IRAS detected QSOs and ULIRGs.  QSOs cluster in the area
$\alpha_{25}^{60} \approx \alpha_{60}^{100} \approx -1$.  Well-known
star-forming QSOs (e.g., Canalizo \& Stockton 2000a) occupy the area
close to the ULIRGs in this FIR color-color diagram.  Indeed
$\alpha_{25}^{60}$ appears a good discriminator: in fact some strongly
star-forming QSOs (genuine PG QSOs!) are too `cool' to classify as AGN
under the De Grijp et al.  (1985) $\alpha_{25}^{60} > -1.5$ criterion. 
Such `cool' QSOs were addressed by Clements (2000 -- see Section~2) and
are also under study by Stockton and collaborators, Sanders and
collaborators, and by our group -- the next section describes this work. 

\begin{figure}[h]

\putplot{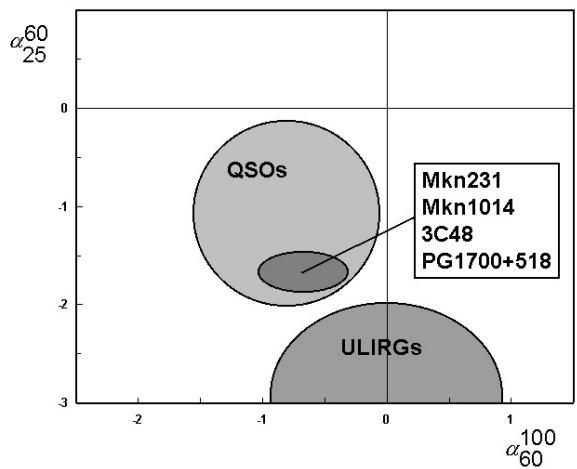}{6cm}{0}{70}{70}{-20}{-187}

\caption{IRAS color--color diagram for quasars and ULIRGs}

\end{figure}

\section{The starburst--AGN symbiosis: source evolution?}

As we have seen in the previous section, the far-infrared spectral index
$\alpha_{25}^{60}$ broadly speaking allows to address the relative
importance of AGN-activity and starburst-activity in -- at least not too
distant -- galaxies.  Attempts have also been made to combine an
infrared spectral index with the relative PAH strength (e.g., Lutz et
al.  1998) and with line ratio's (e.g., Kewley et al.  2001) in such
assessments.  A Groningen-lead study (in a collaborative effort with
researchers in Hawaii, Socorro, Cambridge (UK) and Hertfordshire) has
shown that the FIR SEDs indeed present a powerful tool, particularly
25 and 60\,$\mu$m photometry in combination with sensitive
high-resolution radio imaging. 

Our group has obtained and analysed deep radio (VLA), optical/nearIR
(ESO; La Palma, including Carlsberg Meridian Circle astrometry), as well
as improved IRAS data for a sample of 16 Seyfert galaxies having
$z<0.02$, and 27 radio-quiet PG QSOs having $0.02<z<0.4$.  These
complementary sets of active galaxies span a wide range (3.5 orders of
magnitude) of luminosities.  These are expressed as $L(12\,\mu$m), and
hence predominantly reflect the AGN strength (e.g., Spinoglio \& Malkan
1989).  The radio-imaging, reaching noise levels of $\sim 30\,{\mu}$Jy,
combined with the optical astrometry yields AGN positions to a 3$\sigma$
accuracy of $\sim 0.4$\,arcsec.  This allows assessment of the
star-formation driven radio emission.  The radio data of these -- I
stress -- radio-quiet active galaxies are being combined with
far-infrared photometry, yielding $u$-parameters
log\,$S_{60\mu}/S_{6{\rm cm}}$ (see e.g., Condon \& Broderick 1988),
which permit the determination of the relative contributions of nuclear
activity and star-formation.  Most Seyfert galaxies have infrared
detections (at 25 and 60\,$\mu$m), in contrast to about half of the PG
QSOs.  Also the ratios of the 60\,$\mu$m and blue flux densities were
compiled. 

Figure~5 shows $\alpha^{60}_{25}$ as function of the $u$-parameter. 
Normal star-forming spirals, obeying the radio-FIR correlation, have $u$
values in the range 2.4--3.0.  It is seen that the more powerful
(radio-quiet but {\it not} radio-silent) QSOs have flatter indices than
the Seyferts, and that excess nuclear radio emission (i.e., $u < 2.4$)
leads to warmer dust for both classes. 

\begin{figure}[h]

\putplot{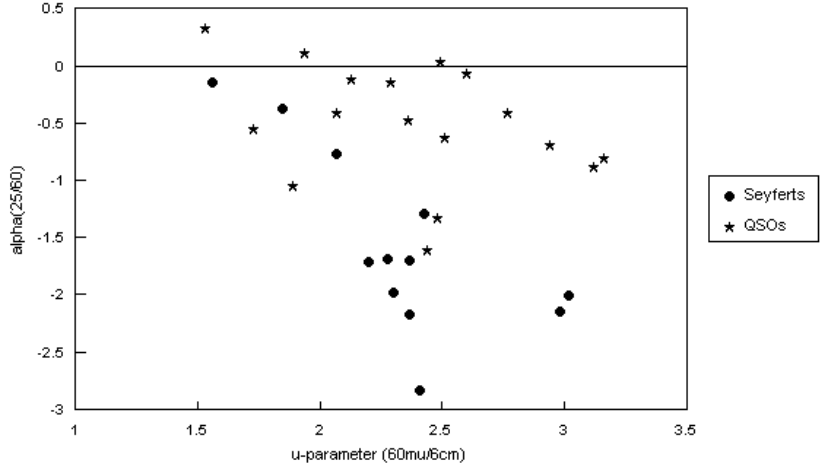}{6.5cm}{0}{100}{100}{-80}{-307}

\caption{Infrared spectral index $\alpha_{25}^{60}$ as function of 
 $u$-parameter log\,$S_{60\mu}/S_{6{\rm cm}}$}

\end{figure}

In order to quantify the heating effect of the AGN luminosity, I plot in
Figure~6 the FIR spectral index as function of the 12\,$\mu$m
luminosity.  The trend of warmer dust with increasing AGN luminosity is
clearly seen.  Examination of the $u$-parameters results in the
distributions as marked with light and dark shading.  Light shading
indicates objects having $u<2.5$, having excess nuclear radio emission. 
Their $60{\mu}/B$ ratios are low.  Dark shading indicates objects with
merely diffuse radio emission, $u \sim 2.7$ and $60{\mu}/B$ ratios
typically a factor 5 -- 10 higher.  Note that particularly for the QSOs
the $B$-band magnitude measures the AGN strength. 

\begin{figure}[h]

\putplot{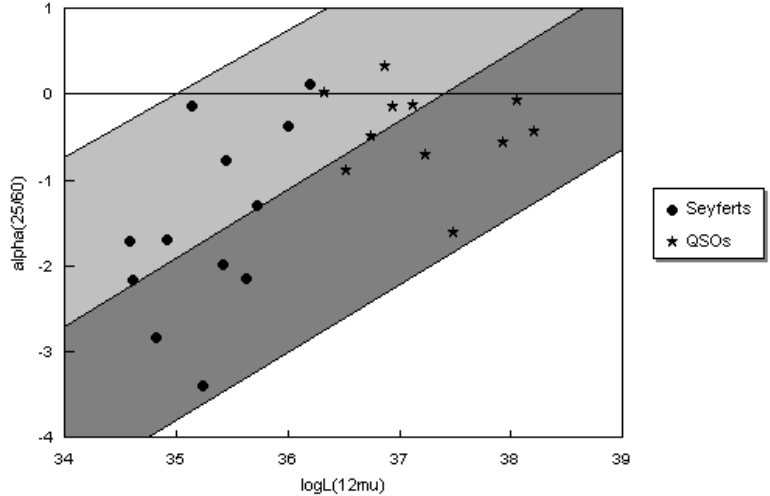}{6cm}{0}{90}{90}{-70}{-270} 

\caption{Infrared spectral index $\alpha_{25}^{60}$ as function of AGN
 luminosity, with star-formation being relatively important (dark grey)
 and unimportant (light grey).}

\end{figure}

On the basis of these measurements I conclude that star-formation plays
a stronger role in the lower half of the diagonal strip than it does in
the upper half.  Could this be an evolutionary effect, during the
lifetime of the active galaxy? Figure~7 is the same plot as Figure~6,
but with the well-known (U)LIRGs Arp\,220, NGC\,6240, Mkn\,231 and
star-forming Seyfert galaxies NGC\,1808, NGC\,7469, NGC\,7552 added. 
These strongly star-forming systems indeed appear in and below the lower
part of the diagonal strip, in line with the picture where the AGN phase
is preceded by a gradually weakening starburst phase.  Note that for the
six star-formers (open symbols) the 12\,$\mu$m luminosity is probably
not the optimal indicator of their FIR luminosity.  Their actual
luminosity points may lie more to the right.  When rotated by
$90^{\circ}$, figure~7 displays luminosity vs.  temperature, and can
thus be considered as the active galaxy equivalent of the classical
Hertzsprung-Russell diagram.  Larger samples and detailed analysis of
the star-formation histories (e.g., Canalizo \& Stockton 2000a, Canalizo
2000) may permit to draw evolutionary tracks.  JCMT observations of the
molecular gas in QSO hosts are currently being analyzed by our group to
test the evolutionary scenario.  Given that roughly half of the AGN
occupy the strip where star-formation is important, the similarity of
the cosmic star-formation history and the quasar space density history
referred to in the introductory section may indeed {\it not} be
coincidental. 

\begin{figure}[h]

\putplot{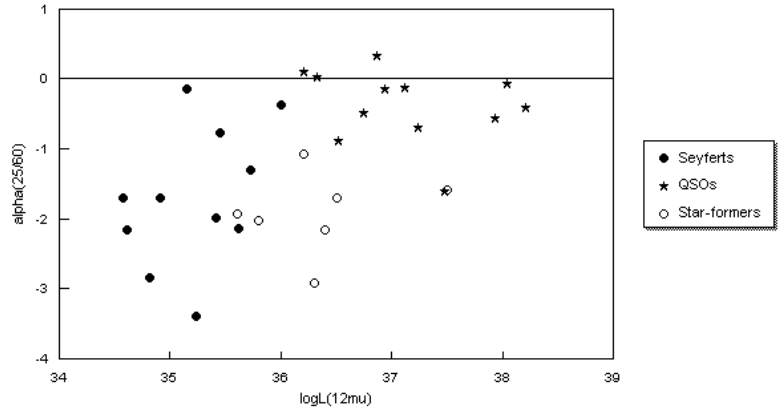}{5.5cm}{0}{100}{100}{-80}{-300}      

\caption{An AGN Hertzsprung-Russell diagram?}

\end{figure}

\section{Summary}

The 60\,$\mu$m luminosity, when normalized with the 25\,$\mu$m (or the
12\,$\mu$m), the blue or the radio luminosity, permits one to assess the
absolute and relative strength of star-formation and nuclear activity in
active galaxies and quasars.  The FIR temperature can be combined with
measures of the bolometric luminosity to yield an intriguing AGN
Hertzsprung-Russell diagram, which among other things suggests that
star-formation plays an important role.  These photometric ratios can be
obtained in a straightforward manner for the faint distant objects to be
measured in large quantities with upcoming space-infrared missions, such
as SIRTF, ASTRO-F and FIRST/Herschel. 

\section{Acknowledgements}
Thanks are due to Dieter Lutz, David Sanders and Ilse van Bemmel, who
allowed me to reproduce figures, and to Jim Higdon who read the
manuscript.  I acknowledge a long collaborative effort with Bob Argyle,
Jeroen Gerritsen, Magiel Janson, Johan Knapen, David Sanders and Dick
Sramek. 

\label{}

\end{document}